\documentclass[secnumarabic,
amssymb, nobibnotes, nofootinbib, aps, showpacs,
tightenlines,
preprint
]{revtex4-1}
\usepackage{amsmath,amssymb,amsthm}
\usepackage{latexsym,graphicx,color,subfigure}
\usepackage[bookmarks=true]{hyperref}
\usepackage{wrapft}
\usepackage{mathrsfs}
\usepackage{tikz}
\usepackage{slashed}
\usepackage{pdfpages}
\usepackage[T1]{fontenc}
\usepackage{amsthm}
\usepackage{amsfonts}
\usepackage{bm}
\usepackage{cancel}
\usepackage{pgffor}
\usepackage[pre-1.03]{feynmf}

\newcommand{\be}{\begin{equation}}
\newcommand{\ee}{\end{equation}} 
\newcommand{\beq}{\begin{eqnarray}}
\newcommand{\eeq}{\end{eqnarray}}

\newcommand{\D}{\mathcal{D}}
\newcommand{\p}{\partial}
\newcommand{\Tr}{{\rm Tr}}

\newcommand{\bea}{\begin{eqnarray}}
\newcommand{\eea}{\end{eqnarray}}
\def\Tr{ \hbox{\rm Tr}}

\def\half{\frac{1}{2}}

\def\half{\frac{1}{2}}
\def\dcfl{\Delta_{\textsc{cfl}}}

\def\SU{\rm{SU}}
\def\U{\rm{U}}

\def \L{\textsc{l}}
\def \R{\textsc{r}}
\def \B{\textsc{b}}
\def \c{\textsc{c}}
\def \F{\textsc{f}}

\def\inv{{-\,1}}
\def\T{{\rm T}}
\def\n{\hat n}
\def \f{f(|\phi|)}

\begin{document}
\setlength{\unitlength}{1mm}
\title{{\bf Coupling between Majorana fermions and 
Nambu-Goldstone bosons 
inside a non-Abelian vortex in dense QCD}}
\author{Chandrasekhar Chatterjee$^{1, a}$, Mattia Cipriani$^{2, b, \dagger}$ and Muneto Nitta$^{1,c}$ }
\email{chandra.chttrj@gmail.com$^{a}$, chandra@phys-h.keio.ac.jp$^{a}$,   mttcipriani@gmail.com$^{b}$, 
\\ nitta(at)phys-h.keio.ac.jp$^{c}$}
\affiliation{$^1$ Department of Physics, and Research and Education Center for Natural Sciences,\\ Keio University,Hiyoshi 4-1-1, Yokohama, Kanagawa 223-8521, Japan}
\affiliation{$^2$ Department of Physics ``E. Fermi'', INFN, University of Pisa, Largo Bruno Pontecorvo 7, Pisa 56127, Italy}
\altaffiliation{Present address: ENEA, FSN-FUSPHY-SAD, C. R. Frascati, via ``E. Fermi'' 45, 00044 Frascati, Italy.}
\date{\today}
\begin{abstract}
Quark matter at high density may exhibit color superconductivity.
As magnetic flux tubes in metallic superconductors, 
color magnetic flux tubes as non-Abelian vortices appear in 
the color-flavor locked phase of high density QCD.
${\mathbb C}P^2$ Nambu-Goldstone bosons 
and Majorana fermions belonging to the triplet 
representation are known to be localized around 
a non-Abelian vortex. 
In this paper, we determine the coupling of these bosonic and fermionic modes
by using the nonlinear realization method.

\end{abstract}
\pacs{ 21.65.Qr, 11.27.+d, 12.38.-t, 25.75.Nq, 11.10.Lm, 11.30.Ly}
\maketitle
\newpage
\section{Introduction}

It is an interesting question what a state of matter becomes 
at extremely high density such as the cores inside neutron stars. 
   At very high densities, the quarks are expected to 
form Cooper pairs to 
show color superconductivity \cite{Alford:1997zt,Alford:1998mk}  
(see Refs.~\cite{Alford:2007xm,Rajagopal:2000wf} 
as a review). At intermediate densities the  up and down quarks  participate in condensation to form
the two-flavor superconducting (2SC) phase.
The 
color-flavor locked (CFL) phase, where all  three flavors(up, down and strange quarks)
participate in condensations, 
is expected to 
be realized 
at asymptotically high densities.  In this phase 
the $\SU(3)_{\c} \times \SU(3)_{\L} 
\times \SU(3)_{\R} $  symmetry  is broken down to
$ \SU(3)_{\c+\L+\R}$ 
\cite{Alford:1998mk,Alford:1999pa,Alford:2007xm}.
 The gap function in the CFL phase 
is given by $\Delta^{\alpha i} 
\propto \epsilon^{\alpha \beta \gamma} \epsilon^{ijk}  
\langle \psi^{\beta j} \psi^{\gamma k} \rangle$, where
$i$, $j$, $k$ are color indices, and $\alpha$, $\beta$,
$\gamma$ are flavor indices. As the magnetic flux tubes can be introduced in 
a type-II metallic superconductor 
 in the presence of a magnetic field, the
color magnetic flux tubes are present stably 
in the CFL phase as well
\cite{Balachandran:2005ev,Nakano:2007dr,Nakano:2008dc,Eto:2009kg};
see Ref.~\cite{Eto:2013hoa} as a review.
In the framework of the Ginzburg-Landau approach \cite{Giannakis:2001wz,Iida:2000ha,Iida:2001pg},  
the detailed profile of the gap  function was 
calculated in Ref.~\cite{Eto:2009kg}.
An important consequence of non-Abelian vortices is that,  
the color-flavor-locked symmetry $\SU(3)_{\c+\L+\R}$ 
of the ground state CFL phase is spontaneously broken down to 
its subgroup $\SU(2)_{\c+\L+\R} \times \U(1)_{\c+\L+\R}$ 
in the presence of a non-Abelian vortex. 
This symmetry breaking generates the Nambu-Goldstone bosons
 around the vortex, which parametrize 
a coset space
$
\mathbb{C}P^{2} \simeq \SU(3)_{\c+\L+\R} 
 / [SU(2)_{\c+\L+\R} \times U(1)_{\c+\L+\R}]
$ \cite{Nakano:2007dq,Nakano:2007dr}.  Accordingly, the vortex 
equations allow a continuous family of degenerate solutions,  corresponding to $\mathbb{C}P^{2}$. 
These modes are known as  the ``internal orientational zero modes''  
and each point of $\mathbb{C}P^{2}$ corresponds to the direction of color flux 
which the non-Abelian vortex carries. 
These  $\mathbb{C}P^{2}$ modes are localized around the vortex and it is possible to   
 construct a two-dimensional $(t, z)$ effective action
on the vortex world sheet by integrating 
the original action over the coordinates $(x, y)$ perpendicular to vortex direction along the $z$-axis \cite{Eto:2009bh,Eto:2009tr}.
The interactions of the localized $\mathbb{C}P^2$ modes
with the gluons \cite{Hirono:2010gq}
and photons \cite{Hirono:2012ki,Vinci:2012mc} in the bulk were studied before. 
When the CFL phase is rotating, first, $\U(1)$ superfluid vortices are created \cite{Forbes:2001gj} 
as liquid helium superfluids. 
Then, the superfluid vortex decays into three color flux tubes
\cite{Alford:2016dco} and they will constitute a non-Abelian vortex lattice. 
A lattice of non-Abelian vortices 
shows a color ferromagnetism
\cite{Kobayashi:2013axa}.
Recently, it has also been shown that 
a non-Abelian vortex has the Aharanov-Bohm phase 
\cite{Chatterjee:2015lbf}.

The above results were derived based on
the Ginzburg-Landau effective theory, 
which  is a macroscopic theory valid only at the 
scale larger than the penetration depth and coherence length and at temperatures 
close to the transition temperature.
In order to figure out the whole structure of 
the non-Abelian vortices in the color superconducting phase, including the region inside their cores, 
it is necessary to  consider fermion dynamics from the  Bogoliubov-de Gennes (BdG) equation.
By solving the BdG equations with the given vortex gap profile, it is possible to find fermion zero 
modes 
in the center of the vortex. 
In the case of a non-Abelian vortex,  
Majorana fermion zero modes belonging to 
the triplet of $\SU(2)_{\rm C+F}$ were found in 
Ref.~\cite{Yasui:2010yw} 
that are also in agreement with 
the index theorem \cite{Fujiwara:2011za}.
One remarkable consequence of such Majorana fermion 
zero modes is that 
vortices obey non-Abelian statistics \cite{Yasui:2010yh} 
as the case of chiral $p$ wave superconductors 
\cite{Ivanov:2001}, 
if vortices are parallel or confined in two-dimensional plane.

Questions arising immediately are whether 
fermion zero modes localized around the vortex
interact with 
$\mathbb{C}P^{2}$ Nambu-Goldstone 
bosons that are also 
 localized around the vortex, and if so, 
how they interact and how one can 
derive the effective action to describe 
their interaction.  
A drawback of the BdG formalism is that 
it is not easy to deal with bosonic modes
such as 
 $\mathbb{C}P^{2}$ modes that we are considering.
To overcome this problem, we use 
the nonlinear realization method 
that was introduced for the first time in particle physics long ago \cite{Weinberg:1968de, Coleman:1969sm,Salam:1969rq}, in the context of chiral symmetry breaking while describing the dynamics  of various interacting pion fields. It is well known today that symmetries play 
a significant role in  determining the low-energy interactions of massless particles. When symmetries  are broken spontaneously by the ground state of the system,  massless Nambu-Goldstone bosons appear in this process. The scattering amplitudes of these Nambu-Goldstone modes are uniquely  determined by certain low-energy theorems\cite{Bando:1987br,Adler:book}. 
The easiest way to understand the implications of low-energy theorems is  to construct a low-energy effective action in terms of nonlinearly transforming Nambu-Goldstone bosons under  the broken symmetry group.  
The generic  framework of nonlinear realization is in general used to construct a low-energy effective action 
\cite{Coleman:1969sm}.

In this paper we use  the nonlinear realization method 
to construct the effective action to describe 
the coupling between the triplet Majorana fermions and 
${\mathbb C}P^2$ Nambu-Goldstone bosons inside a non-Abelian vortex 
in the CFL phase of dense QCD. 

       The  paper is organized as followes. In Sec.~\ref{CFL}, we discuss the fermion zero mode solutions of the BdG equations  that transform 
under the low-energy group in the presence of a generic vortex profile function 
(see \cite{Fujiwara:2011za, Yasui:2010yw} for a more detailed derivation). We divide  the  Sec.~\ref{NLR} in two subsections. In the first part, Sec.~\ref{NLR1},
we recall the general framework of nonlinear realization of a compact, connected, semisimple Lie group $G$ which is spontaneously broken down to  a specified subgroup $H$ and in Sec.~\ref{NLR2} we apply the nonlinear realization method to the two-dimensional effective action 
for the fermion zero modes in the presence of the $\mathbb{C}P^2$ bosonic zero modes.
Section~\ref{sec:summary} is devoted to summary and discussion. 

\section{ \label{CFL}Fermionic zero modes and effective action}

In this paper, 
our intention is to study the nonlinear realization due to  the action of the color-flavor locked symmetry $\SU(3)_{\c+\F}$($\SU(3)_{\c+\L+\R}$) on the low-energy massless fields in the CFL phase of dense QCD  and derive the effective action of these modes. 
To do so, we start with the BdG equation and it can be determined  from the  generic Hamiltonian with a given pairing gap($\Delta$) of superconductivity,
\begin{eqnarray}
\label{H}
{\cal H} \, = \, \bar\Psi_i^\alpha\, \left(\hat{\cal H}_0\,\delta_{ij} \, \delta^{\alpha\beta}  \, + \, \tilde \Delta^{\alpha\beta}_{ij} \, \right) \, \Psi_j^\beta
\end{eqnarray}
 where $ i,j = \{r, g, b\}, \alpha, \beta = \{u, d, s\}$.  $ \hat{\cal H}_0$ and $\tilde \Delta^{\alpha\beta}_{ij} $ are defined as
 \begin{eqnarray}
 \hat{\cal H}_0 = 
\left(
\begin{array}{cc}
 -i \vec\gamma \cdot \vec\nabla - \gamma^0 \mu & 0  \\
 0  &    -i \vec\gamma \cdot \vec\nabla  +  \gamma^0 \mu\\
\end{array}
\right), \qquad
\tilde \Delta^{\alpha\beta}_{ij} = 
\left(
\begin{array}{cc}
0  &  \Delta^{\alpha\beta}_{ij} \, \gamma^5  \\
 - {\Delta^*}^{\alpha\beta}_{ij} \,  \gamma^5 &   0  \end{array}
\right).
\end{eqnarray}
Here   $\Delta^{\alpha\beta}_{ij}$ is the generic gap function,  and $\Psi$ is the fermion written in the Nambu-Gor'kov basis  as
\begin{eqnarray}
\Psi = 
\left(
\begin{array}{c}
 \psi  \\
  \psi_c   
\end{array}
\right), \qquad \psi_c \, = \,  e^{i\eta_c} i \gamma^2 \,  \psi^*, 
\end{eqnarray}
where $\eta_c$ is an arbitrary phase  and $\psi$ is defined as a $3\times 3$ matrix whose elements consist of quarks with color and flavor indices,
\begin{eqnarray}
\label{psiquarks}
\psi_i^\alpha=(\psi)_{\alpha i}= 
\left(
\begin{array}{ccc}
u_r \,& u_g\, & u_b\\
d_r \,& d_g \,& d_b\\
s_r \,& s_g \,& s_b
\end{array}
\right).
\end{eqnarray}
  In this paper we work in the Nambu-Gor'kov basis where  $\psi$ and $\psi_c$ are treated as particle and hole.
 This Hamiltonian density (\ref{H}) leads us to the BdG eigenvalue equation, 
 \begin{eqnarray}
 \label{BDG1}
{\cal H}\Psi = {\cal E} \Psi .
\end{eqnarray}
We assume the form of the gap function to be
\begin{eqnarray}
\Delta^{\alpha\beta}_{ij} \, = \, \epsilon^{\alpha\beta\gamma}\,\epsilon_{ijk}\, \Delta_\gamma^k.
\end{eqnarray}
The transformation properties of the gap function under the action of
$(e^{i\theta},U_{\c},U_{\F}) \in 
{\U(1)}_{\B} \times {\SU(3)}_{\c}\times {\SU(3)}_{\F}$
can be expressed  as
\begin{eqnarray}
&& \Delta_\gamma^k \,\,\,
 \to \,\,\, e^{i\theta} \, (U_{\c})_\gamma^\alpha \,\,
     \Delta_\alpha^i \,\, (U_{\F}^T)_i^k .
\end{eqnarray}
At the CFL phase ground state, the gap function takes the form of 
\begin{alignat}1
\Delta_\gamma^k\,\, = \,\, \dcfl \, {\bf 1_3} 
\end{alignat}
 where $\dcfl$ is a constant.
This gap function breaks the ${\rm U(1)}_{\B} \times {\rm SU(3)}_{\c}\times {\rm SU(3)}_{\F}$ symmetry and only the  diagonal symmetry group, the so-called color-flavor locking 
symmetry ${\rm SU}(3)_{\c+\F }$, is preserved. 

In the presence of a vortex, the situation changes, because the gap function  no longer remains constant.
 For the minimal winding non-Abelian vortices, $\Delta_\gamma^k$ is given by 
\begin{eqnarray}
\Delta_\gamma^k
=\mbox{diag}\left(\Delta_0, \Delta_0,\Delta_1\right) ,
\label{eq:na-vortex}
\end{eqnarray}
where $\Delta_1(r,\theta)=|\Delta_1|e^{i\theta}$ with 
the boundary conditions 
$\Delta_0(\infty)=|\Delta_1(\infty)|=\Delta_{\rm CFL}$ and
$|\Delta_0(0)|' = |\Delta_1(0)|=0$. A detailed profile of $\Delta$ can be derived numerically by solving the BdG equation and the gap equation self-consistently,
 but it has yet to be done. 
This vortex carries  a color magnetic flux  confined inside its core (we have 
  neglected it  in the BdG equation).
The CFL symmetry ${\rm SU}(3)_{\c+\F}$ is now broken 
  spontaneously 
into ${\rm SU}(2)_{\c+\F}\times {\rm U}(1)_{\c+\F}$ inside the vortex core.
This generates Nambu-Goldstone bosonic zero modes which 
parametrize 
the ${\mathbb C}P^2 \simeq {\rm SU}(3)_{\c+\F}/[{\rm SU}(2)_{\c+\F}
\times {\rm U}(1)_{\c+\F}]$  moduli space.
The configuration in Eq.~(\ref{eq:na-vortex}) 
corresponds to a particular point 
on ${\mathbb C}P^2$. 
These bosonic modes are gapped if one takes into account
either the presence of the strange quark mass 
\cite{Eto:2009tr} or 
the nonperturbative quantum effect in 
the vortex world-sheet theory \cite{Gorsky:2011hd,Eto:2011mk}.

  In this section we are interested in writing down low-energy effective action in terms of fermion zero modes  that transform under the unbroken ${\rm SU(2)_{\c+\F}}$, discussed above.  The explicit form of the zero modes can be found by solving the BdG equation (\ref{BDG1}). To do so let us expand $\Psi$ in the ${\rm U(3)}$ generators as
  \begin{eqnarray}
\Psi_{\alpha i} = \psi^a T^a_{\alpha i} &,& \qquad \Tr( T^a T^b) = \delta^{ab},
\nonumber \\
T^9 = \frac{1}{\sqrt 3} {\bf 1_3}&,& \qquad T^a =  \frac{1}{\sqrt 2} \lambda^a, \qquad \{a, b\} = \{ 1, 2\cdots, 8\}
\end{eqnarray}
where $\lambda^a$ are the Gell-Mann matrices.
Now the Hamiltonian (\ref{H}) can be  rewritten in this basis as 
\begin{eqnarray}
{\cal H}\,\,  = \,\,  \bar\Psi^a\, \left(\hat{\cal H}_0\, \delta_{ab} \, + \, \Theta_{ab} \, \right)\, \Psi^b,\qquad \{a, b\}  = [1, 2\cdots, 9]
\end{eqnarray}
where
\begin{eqnarray}
\Theta_{ab} &=& T^a_{i\alpha} \, \tilde\Delta^{\alpha\beta}_{ij} \, T^b_{\beta j} \nonumber \\
&=& 
\left(
\begin{array}{ccccccccc}
 -  \tilde\Delta_1& & &&&&&&\\
&  \tilde\Delta_1 &  &&&&&&\\
 && -  \tilde\Delta_1  &&&&&&\\
&&& - \tilde\Delta_0& &&&&\\
&&&   & \tilde\Delta_0 &&&&\\
&&&&& 0 \, \,& &&\\
&&&&&&0 \, \,  &&\\
&&&&&&& \frac{1}{3}( \tilde\Delta_1 -  4\tilde\Delta_0)&\frac{\sqrt 2}{3}(\tilde \Delta_1 - \tilde \Delta_0)  \\
&&&&&&&\frac{\sqrt 2}{3}(\tilde \Delta_1 - \tilde \Delta_0)  & \frac{2}{3}(2 \tilde\Delta_0 + \tilde\Delta_1) \\
\end{array}
\right)
\label{HamTot}
\end{eqnarray}
and the equations for the triplet states can be written as
\begin{eqnarray}
\label{tripleteq}
\left(\hat{\cal H}_0 -  \tilde\Delta_1\right)\Psi^1 = 0,\qquad \left(\hat{\cal H}_0 +  \tilde\Delta_1\right)\Psi^2 = 0,\qquad  \left(\hat{\cal H}_0 -  \tilde\Delta_1\right)\Psi^3 = 0.
\end{eqnarray}
The equations for singlet states are
\begin{eqnarray}
\left(
\begin{array}{cc}
 \hat{\cal H}_0  +  \frac{1}{3}( \tilde\Delta_1 -  4\tilde\Delta_0)&\frac{\sqrt 2}{3}(\tilde \Delta_1 - \tilde \Delta_0)  \\
\frac{\sqrt 2}{3}(\tilde \Delta_1 - \tilde \Delta_0)  & \hat{\cal H}_0  + \frac{2}{3}(2 \tilde\Delta_0 + \tilde\Delta_1)
\end{array}
\right)\left(
\begin{array}{c}
 \Psi^8  \\
  \Psi^9 
\end{array}
\right) = 0.
\end{eqnarray}
Following Eq.~(\ref{psiquarks}) the triplet can be expressed as
\begin{eqnarray}
\psi^1 = \frac{d_r + u_g}{\sqrt 2}, \qquad \psi^2 =  \frac{d_r - u_g}{i\sqrt 2}, \qquad  \psi^3 = \frac{u_r - d_g}{\sqrt 2},
\end{eqnarray}
while the singlets are identified as
\begin{eqnarray}
\psi^8 =  \frac{u_r + d_g - 2 s_b}{\sqrt 6} , \qquad \psi^9 =  \frac{u_r + d_g + s_b}{\sqrt 3}.
\end{eqnarray}
As  was shown in Refs.~\cite{Yasui:2010yw,Fujiwara:2011za}, 
the singlet mode is non-normalizable. Here we are only concerned with the triplet zero modes as low-energy excitations. The  triplet  zero mode quarks are 
analytically found by solving Eq.~(\ref{tripleteq}) by expressing fermions as
\begin{eqnarray}
\Psi^{(1)}_{{R}}(r,\theta)= C_{1}
\left(
\begin{array}{c}
 \varphi_{{R}}(r,\theta) \\
 \eta_{{R}}(r,\theta)
\end{array}
\right), \ 
\Psi^{(2)}_{{R}}(r,\theta) = C_{2}
\left(
\begin{array}{c}
 \varphi_{{R}}(r,\theta) \\
 -\eta_{{R}}(r,\theta)
\end{array}
\right), \ 
\Psi^{(3)}_{{R}}(r,\theta) = C_{3}
\left(
\begin{array}{c}
 \varphi_{{R}}(r,\theta) \\
 \eta_{{R}}(r,\theta)
\end{array}
\right)\nonumber
\\
\Psi^{(1)}_{L}(r,\theta)= C'_{1}
\left(
\begin{array}{c}
 \varphi_{L}(r,\theta) \\
 \eta_{{L}}(r,\theta)
\end{array}
\right), \ 
\Psi^{(2)}_{{L}}(r,\theta) = C'_{2}
\left(
\begin{array}{c}
 \varphi_{{L}}(r,\theta) \\
 -\eta_{{L}}(r,\theta)
\end{array}
\right), \ 
\Psi^{(3)}_{{L}}(r,\theta) = C'_{3}
\left(
\begin{array}{c}
 \varphi_{{L}}(r,\theta) \\
 \eta_{{L}}(r,\theta)
\end{array}
\right) \nonumber\\
\end{eqnarray}
where $C_i$ and $C'_i$ are normalization constants. The explicit solutions of $\varphi_{R/L}$(particle) and $\eta_{R/L}$(hole) 
 can be  found to be
\begin{eqnarray}
 \varphi_{{R}}(r,\theta) 
= {\rm e}^{-\int_{0}^{r} |\Delta_{1}(r')|\mbox{d}r'}
\left(
\begin{array}{c}
 J_{0}(\mu r) \\
 i J_{1}(\mu r)\, {\rm e}^{i\theta}
\end{array}
\right), && \hspace{0.5em}
 \eta_{{R}}(r,\theta) = {\rm e}^{-\int_{0}^{r} |\Delta_{1}(r')|\mbox{d}r'}
\left(
\begin{array}{c}
 J_{1}(\mu r)\, {\rm e}^{-i\theta} \\
- i J_{0}(\mu r) \\
\end{array}
\right) \nonumber\\
\varphi_L(r,\theta) = {\rm e}^{-\int_{0}^{r} |\Delta_{1}(r')|\mbox{d}r'}
\left(
\begin{array}{c}
 J_0(\mu r) \\
  -i J_1(\mu r) e^{i\theta} 
\end{array}
\right),&&  \hspace{0.5em}
 \eta_{L}(r,\theta) = {\rm e}^{-\int_{0}^{r} |\Delta_{1}(r')|\mbox{d}r'}
\left(
\begin{array}{c}
- J_1(\mu r) e^{- i\theta}  \\
 - i  J_0(\mu r)
\end{array}
\right),
\label{tripletZMs}\nonumber\\
\end{eqnarray}
for the generic shape of the vortex profile $|\Delta_{1}(r)|$, where 
 $J_{0,1}$ are the Bessel functions. It should be noticed that  the zero mode solutions written above  satisfy the following  ``Majorana condition'' 
 \begin{eqnarray}
\Psi_{R/L} = \kappa \gamma^2 \Psi_{R/L}^*
\end{eqnarray}
where $\kappa = \pm 1$ for right and left modes, respectively.

   As we have seen that the zero modes are eigenstates of the operator $\left(\hat{\cal H}_0\delta_{ab} + \Theta_{ab}\right)$, so the effective action that contains only zero modes may be written as 
\begin{eqnarray}
{\cal L}_{\rm eff} = \int d^2x\,\,\, \bar \Psi(t, z, x, y)(-i \gamma^I\p_I)\Psi(t, z, x, y), \qquad I = \{0, 3\}.
\end{eqnarray}
 The $z$  dependence enters here in a factorized way:

 \begin{eqnarray}
\Psi_L^a(t, z, x, y) = \chi_L^a(t, z) {\Psi_0}_L^a(x, y), \qquad \Psi_R^a(t, z, x, y) = \chi_R^a(t, z) {\Psi_0}_R^a(x, y).
\end{eqnarray}
where $a = \{ 1, 2,3\}$(no summation over $a$). The two-dimensional spinors ${\chi}^a(t, z)$ are defined by
\begin{eqnarray}
{\chi}^a(t, z) = 
\left(
\begin{array}{ccc}
 {\chi}_L^a(t, z)  \\
 {\chi}_R^a(t, z) \end{array}
\right).
\end{eqnarray}

Using the normalization condition $\int d^2x\,\,\, \Psi_0^\dagger(x, y)\Psi_0(x, y) = 1$, the effective action can be derived in terms of the two-dimensional spinor
${\chi}^a(t, z)$ as
\begin{eqnarray}
{\cal L}_{\rm eff} &=& -i \chi^\dagger_a\left( \dot\chi^a - v(\mu, \Delta)\p_z \chi^a\right) \nonumber \\
&=& -i \Tr \,\chi^\dagger\left( \dot\chi - v(\mu, \Delta)\p_z \chi\right), \qquad \chi = \chi^a \tau^a
\end{eqnarray}
where $v(\mu, \Delta)$ is the velocity defined by
\begin{eqnarray}
v_{L/R}(\mu, \Delta) = \int d^2x\, {\Psi_0}_{L/R}^\dagger  \gamma^0\gamma^3{\Psi_0}_{L/R}.
\end{eqnarray}
This can be expanded as 
\begin{eqnarray}
v_{L/R}(\mu, \Delta) = \int d^2x\,\, \big[ \psi_{L/R}^\dagger(r, \theta)\sigma^3\psi_{L/R}(r, \theta) - {\psi_{c}}_{L/R}^\dagger(r, \theta) \sigma^3 {\psi_c}_{L/R}(r, \theta)\big]
\end{eqnarray}
where $\psi$ and $\psi_c$ are particle and hole solutions, respectively. The exact value of $v_{L/R}(\mu, \Delta)$ depends on the exact profile of the vortex. This can be noted if we write the  $v_{L/R}(\mu, \Delta)$ explicitly as
\begin{eqnarray}
v_{\rm fermi} \equiv
v_{L}(\mu, \Delta) =  
v_{R}(\mu, \Delta) = \int d^2x\,\, \big[ \varphi^\dagger(r, \theta)\sigma^3\varphi(r, \theta) - \eta^\dagger(r, \theta) \sigma^3 \eta(r, \theta)\big]. \label{eq:vfermi}
\end{eqnarray}

The fermion zero modes shown above 
are excitations around the  particular vortex 
configuration in Eq.~(\ref{eq:na-vortex}). 
If we consider a configuration having the winding number 
in the second or third diagonal component, 
we have fermion zero modes in a different block.
They all correspond to the particular points 
of the ${\mathbb C}P^2$ moduli space. 
The general vortex configurations correspond to 
 the ${\mathbb C}P^2$ moduli space, 
and by considering them all together one can 
discuss the interaction between the fermion zero modes 
and 
 the ${\mathbb C}P^2$ Nambu-Goldstone bosons,
which is the main purpose of this paper 
discussed in the next section.

\section{\label{NLR}Nonlinear Realization in the CFL phase}
In the first subsection, we would like to summarize  briefly the nonlinear  realization of a compact, connected, semisimple Lie group $G$ which is spontaneously broken down to  a specified subgroup $H$.  Now let us consider a situation where  we can introduce fields transforming  linearly only  under the unbroken subgroup $H$ of the full group $G$. In this system,  it is  possible to classify all possible nonlinear realizations of the symmetry group $G$ that become linear  under the action of the unbroken subgroup $H$ \cite{Coleman:1969sm}. In the second subsection we analyze the nonlinear realization for the  CFL phase  where $\SU(3)_{\c+\F}$ symmetry is spontaneously broken down to $\U(1) \times \SU(2)$ by the presence of a vortex and derive the effective action in terms of a Nambu-Goldstone mode and fermion zero modes.

\subsection{\label{NLR1}Nonlinear realization}
 Let us introduce a field $\Psi$ which transforms under  a linear(preferably irreducible) representation of $G$ as
\begin{eqnarray}
\Psi' = D(g)\Psi,\qquad g\in G.
\end{eqnarray}
Now let us suppose that the system goes through a spontaneous symmetry breaking  that leaves $H$, a continuous subgroup of $G$, unbroken. We also assume that the basis has been chosen in order to represent $D(h)$($h \in H$) in a block diagonal form. To derive a nonlinear realization, we may define a ``\cal{reducing matrix},'' ${L_\phi}_{\alpha\beta}$, the elements of which are field variables (Nambu-Goldstone bosons). The properties of $L_\phi$ can be summed up as follows:
\begin{itemize}
\item The matrix $L_\phi$ belongs to the group $G$. For any finite-dimensional representation representation $D(g)$, the $D(L_\phi)$ should be well defined. 
\item The matrix $L_\phi$ transforms as
\begin{eqnarray}
\label{Ltrans}
L'_\phi = g\, L_\phi \, h^\inv, \qquad g\,\in G, 
\quad h\in H.
\end{eqnarray}
i.e. the columns of the reducing matrix transform among themselves according to some representation of $H$. It follows from Eq.~(\ref{Ltrans}) that for any finite-dimensional representation, $D(L_\phi)$ transforms as 
\begin{eqnarray}
D(L_\phi) \,\, \rightarrow \,\, D(g L_\phi h^\inv) = D(g)D(L_\phi) D(h^{-1}).
\end{eqnarray}

\item The subgroup $H$ operates on   the reducing matrix  $L_\phi$ linearly as
\begin{eqnarray}
L'_\phi = h L_\phi h^\inv, \qquad h\in H.
\end{eqnarray}
\end{itemize}
These properties of the reducing matrix help us to represent the nonlinear realization from the linear one by defining a new field,
\begin{eqnarray}
\psi = D(L_\phi^\inv)\Psi.
\end{eqnarray}
The transformation properties of $\psi$ can be derived as
\begin{eqnarray}
\psi' &=& D\left(h_{\phi, g}\right) D(L_\phi^\inv) D(g) D(g^\inv)\Psi
= D\left(h_{\phi, g}\right)\psi.
\end{eqnarray}
Here $h_{\phi, g}$ in general has a nonlinear structure; 
that is, it depends upon the fields ($\phi$) that parametrize $L_\phi$. However, 
the derivative of $\psi$ does not transform covariantly and this problem can be solved by introducing a covariant derivative as
\begin{eqnarray}
\label{covariantD}
\D_\mu \psi = \p_\mu\psi +  i e_\mu^a D(h_a)\psi,
\end{eqnarray}
where  $  e_\mu^a h_a$  would behave as a gauge field and it is related to the Maurer-Cartan one-form as
\begin{eqnarray}
\label{MCdecomposition}
-i L^\inv_\phi \p_\mu L_\phi = e_\mu^a h_a + e_\mu^\alpha  T_a,
\end{eqnarray}
where $h^a$ and $T^a$ are the generators of $H$ and $G/H$, respectively.

   The transformation properties of the covariant derivative, defined above, can be derived as follows:
\begin{eqnarray}
\D_\mu \psi' &=& \D_\mu \left(D(h_{\phi, g})D(L_\phi^\inv)\Psi \right)
\nonumber\\
&=& \p_\mu \left(D(h_{\phi, g})D(L_\phi^\inv)\Psi \right) + i  e_\mu^a D(h_a) \left(D(h_{\phi, g})D(L_\phi^\inv)\Psi \right)
\nonumber\\
&=& D(h_{\phi, g})\left( \p_\mu + i {e'_\mu}^a D(h_a)\right)\psi,
\end{eqnarray}
where ${e'_\mu}^a D(h_a)$ is defined by the inhomogeneous gauge transformation as 
\begin{eqnarray}
{e'_\mu}^a D(h_a) = {e_\mu}^a D(h_{\phi, g}^\inv)  D(h_a) D(h_{\phi, g}) -i D(h_{\phi, g}^\inv)\p_\mu D(h_{\phi, g}).
\end{eqnarray}

\subsection{\label{NLR2}Nonlinear realization inside a non-Abelian vortex}

In this subsection we demonstrate the nonlinear realization for the case of fermionic zero modes, discussed in Sec.~\ref{CFL} and derive the two-dimensional effective action.  We consider here the nonlinear realization of   the broken $\SU(3)_{\c+\F}$($G$) symmetry which becomes linear under  unbroken $\SU(2)_{\c+\F}\times\U(1)_{\c+\F}$($H$) color-flavor transformation. This symmetry breaking generates Nambu-Goldstone modes which parametrize the coset space $G/H \simeq {\mathbb C}P^2$ . According to Sec.~\ref{NLR1}, the nonlinear realization introduces covariant derivatives in the effective action. So the new effective action describe the interaction between the fermionic zero modes and ${\mathbb C}P^2$ Nambu-Goldstone bosons via the covariant derivative.  In order to establish the nonlinear realization in this case, we first need to construct the reducing matrix ${L_\phi}_{i\alpha} \in \SU(3)_{\c+\F}$.

   For the purpose  of calculation we choose the basis such that the unbroken  $\SU(2)_{\c+\F}$ subgroup lives in the upper left $2\times 2$ block of a $3\times 3$ matrix of the full symmetry group $\SU(3)_{\c+\F}$.  The fermionic zero mode $\chi$ , as discussed in Sec.~\ref{CFL}, transforms as the adjoint representation of the symmetry group  $\SU(2)_{\c+\F}$.  So the action of  $\SU(3)_{\c+\F}$ on $\chi^a$, which transforms linearly under the unbroken $\SU(2)_{\c+\F}$, would generate a nonlinear realization by introducing the ${\mathbb C}P^2$ Nambu-Goldstone modes $\phi$ into the effective action. 
     
 As described in Refs.~\cite{Coleman:1969sm, Kibble:1967sv}, the simplest way to parametrize the reducing matrix is to write it as  
  \begin{eqnarray}
U(\phi) = e^{i \phi\cdot \T}, \qquad \mbox{where} \,\,\,\,\T \in G/H.
\end{eqnarray}
    Following this   we choose the form of the reducing matrix to be
\begin{eqnarray}
&& U(\phi) = e^{i f(|\phi|) \rm A}, \qquad
\end{eqnarray}
where ${\rm A}, \hat n$ and $ \f$  are defined as
\begin{eqnarray}
{\rm A} = \left(
\begin{array}{cc}
\, {\bf 0_2}  &\,\,\,\, \hat n  \,  \\
\, \hat n ^\dagger &\,\,\,\, 0 \,  
 \end{array}
\right), \qquad
\n =  \frac{1}{\sqrt{|\phi^1|^2 + |\phi^2|^2}} \left(
\begin{array}{c}
  \phi^1  \\
\phi^2  
\end{array}
\right),\qquad  \f = \arctan |\phi|.
\end{eqnarray}

The explicit expression  for the reducing matrix $U$ can be written as
\begin{eqnarray}
U 
&=&\left(
\begin{array}{cc}
\, U_2  &\qquad i U^-\,\\
\,i U^+   &\qquad   U_0\,
\end{array}
\right)  
= {\bf 1_3} - 2 \sin^2 \frac{f}{2}\, {\rm A}^2 + i {\sin} f\,  {\rm A}.
\end{eqnarray}
In the above equation, $U_2$ is a $2\times2$  matrix, $U^-$ is a column, $ U^+ = { U^-}^\dagger$ and  $U_0$ is a number. 
 The components of the unitary matrix $U$ are related by the eigenvalue equation,
\begin{eqnarray}
U_2 U^- = U_0 U^-.
\end{eqnarray}
The explicit expressions of the components  can be written as
\begin{eqnarray}
{U}_2  = {\bf 1_2} -  2 \sin^2 \frac{\f}{2} \,  \n\,\n^\dagger,&&\qquad {U^-} = \sin\f \,\n \nonumber\\  {U^+}  = \sin\f \,\n^\dagger,\qquad\qquad&& \qquad \,{U_0} = \cos\f.
\label{U}
\end{eqnarray}
The homogeneous coordinates of the $\mathbb{C}P^2$ coset space can be expressed in terms of the above coordinates as
\begin{eqnarray}
\label{CP2homo}
 Z = 
\left(
\begin{array}{ccc}
 i U^-  \\
  U_0   
\end{array}
\right) = \left(
\begin{array}{ccc}
 i \sin\f \n  \\
 \cos\f
\end{array}
\right) = \frac{1}{\sqrt{1 + |\phi|^2}} \left(
\begin{array}{ccc}
 i \phi_1  \\
 i\phi_2\\
 1
\end{array}
\right).
\end{eqnarray}

 The covariant derivative can be determined by the Maurer-Cartan one-form,
 \begin{eqnarray}
-i U^\dagger \p_I U = -i \left(
\begin{array}{ccc}
U_2 \p_I U_2 \, +\, U^- \p_I U^+ \qquad&  i  U_2 \p_I  U^- \,- \,i  U^- \p_I U_0 \\
- i U^+\p_I U_2 \, + \, i  U_0 \p_I U^+ \qquad &  U_0\p_I U_0 \,+\,   U^+\p_I \U^-
\end{array}
\right),
\label{MC2}
\end{eqnarray}
where $I = \{0, 3\}$. This matrix belongs to the $\SU(3)$ Lie algebra as described in 
Eq.~(\ref{MCdecomposition}).  So, to construct the covariant derivative  defined in Eq.~(\ref{covariantD}), we need to  separate out the
part which belongs to the  $\SU(2)$ subalgebra. Following the above equation (\ref{MC2}) and using the parametrization of
Eq.~(\ref{U}), we can express $e_I^a(\phi)\, \tau^a$ as
\begin{eqnarray}
\label{Aapprox}
A_I(\phi) &=& e_I^a(\phi)\, \tau^a \nonumber \\
&=& -i \bigg[( U_2 \p_I U_2 \,   + \,  U^- \p_I U^+) - \half\Tr( U_2 \p_I U_2 \,   + \,   U^- \p_I U^+){\bf 1_2}\bigg]\nonumber \\
&=& -i \bigg[\left\{1 - \cos \f\right\}\,\bigg\{\n\,\p_I\n^\dagger - (\p_I\n)\,\n^\dagger +  [1 - \cos \f ]\,(\n^\dagger \p_I\n)\,  \n\,\n^\dagger \bigg\}\nonumber\\ && + \half \sin^2 \f    \,\,\,\n^\dagger(\p_I\n){\bf 1_2} \bigg].
\end{eqnarray}
For small perturbations in the $\mathbb{C}P^2$ space  the approximate form of $A_I(\phi)$ becomes 
\begin{eqnarray}
A_I(\phi) \simeq 
\left(
\begin{array}{ccc}
\Im [\phi_1^* \p_I \phi_1 -   \phi_2^* \p_I \phi_2] &   -i [ \phi_1 \p_I \phi_2^*  - \phi_2^* \p_I \phi_1]   \\
-i [ \phi_2 \p_I \phi_1^*  - \phi_1^* \p_I \phi_2]   &  - \Im [\phi_1^* \p_I \phi_1 -  \phi_2^* \p_I \phi_2]  \\ 
\end{array}
\right).
\end{eqnarray}
So the covariant derivative can be written as 
\begin{eqnarray}
&&\D_I\chi(t,z) = \p_I\chi(t, z) +  i [A_I(t, z), \chi(t,z)]. \nonumber \\
&&\qquad \chi(t, z) = \chi^a(t, z) \tau^a,\qquad \tau^a \in su(2), 
\end{eqnarray}
where $\chi^a$ are the triplet two-component fermionic zero modes, as described above. The final form of the fermionic effective action can be expressed as
\begin{eqnarray}
\label{LEff}
{\cal L}_{\rm fermi} 
&=& -i \Tr \,\left[\chi^\dagger(t, z)\left\{ \D_0\chi(t, z) - v(\mu, \Delta)\D_z \chi(t, z)\right\}\right] .
\end{eqnarray}
Here $v$ is found to be the same for all three triplet zero modes, and in a matrix form it can be written  from Eq.~(\ref{eq:vfermi}) as
\begin{eqnarray}
v = v_{\rm fermi}\, \sigma_3.
\end{eqnarray}

Following Eqs.~(\ref{CP2homo}) and (\ref{MC2}) we can write down the bosonic $\mathbb{C}P^2$ effective action as
\begin{eqnarray}
\label{cp2eff}
{\cal L}_{\mathbb{C}P^2} \,\, = \,\,C_0  \{{\p_0 Z^\dagger} {\p_0 Z} \,\,+\,\, (Z^\dagger \p_0 Z) Z^\dagger {\p_0 Z}\} + C_3 \{{\p_3 Z^\dagger} {\p_3 Z} \,\,+\,\, (Z^\dagger \p_3 Z) Z^\dagger {\p_3 Z}\}.
\end{eqnarray}
where $Z^\dagger Z = 1$ as defined in Eq.~(\ref{CP2homo}) and  the coefficients $C_0$ and $C_3$ are derived in  \cite{Eto:2013hoa, Nakano:2007dq, Eto:2009bh} by using the explicit form of the vortex profile functions in the Ginzburg-Landau theory.

     In order to see the interaction between 
fermionic and bosonic modes more explicitly,
let us reform the effective action.  Let us first rescale the space-time to remove $v_{\rm fermi}$ by setting
\begin{eqnarray}
t \rightarrow t,\qquad z \rightarrow v_{ \rm fermi}\, z.
\end{eqnarray}
This  changes the covariant derivative as
\begin{eqnarray}
v_{\rm fermi}\,\D_z \rightarrow \D_z.
\end{eqnarray}

 We can also define 
  effective  two-dimensional gamma matrices as
 \begin{eqnarray}
\Gamma^t = 
\left(
\begin{array}{ccc}
0  & 1  \\
 1 &   0   
\end{array}
\right), \qquad \Gamma^z = 
\left(
\begin{array}{ccc}
0  & -1  \\
 1 &   0   
\end{array}
\right).
\end{eqnarray}
Then, the effective action can be 
rewritten in a Lorentz invariant way as
\begin{eqnarray}
{\cal L}_{\rm eff} 
&=& -i \Tr \,\bigg[\,\bar \chi(t, z)\, \slashed{\D}\,\chi(t, z)\,\bigg],\qquad \slashed D = \Gamma^I \D_I.
\end{eqnarray}
However this rescaling does not make the theory really Lorentz invariant because this  rescaling is not consistent with the rescaling to make the $\mathbb{C}P^2$ action (\ref{cp2eff})  effectively Lorentz invariant. 

The explicit form of the interaction can be expressed in terms of the components of the fields as
\begin{eqnarray}
\Tr\bigg[\bar\chi\,\,[\slashed A(t, z), \chi(t,z)]\bigg] = i \epsilon^{abc} \bar\chi^a {\slashed A}^b \chi^c, \qquad \slashed A = \Gamma^I A_I
\end{eqnarray}
where the components of $\slashed A$ can be expressed from Eq.~(\ref{Aapprox}) as
\begin{eqnarray}
{\slashed A}^1 = \Im [ \phi_1  \slashed\p \phi_2^*  - \phi_2^*  \slashed\p \phi_1], \quad
{\slashed A}^2 = \Re [ \phi_1  \slashed\p \phi_2^*  - \phi_2^*  \slashed\p \phi_1], \quad
{\slashed A}^3 = \Im [\phi_1^* \slashed\p \phi_1 -   \phi_2^* \slashed\p \phi_2].  
\end{eqnarray}


\begin{figure}[t]
\begin{center}
\includegraphics[width=5cm]{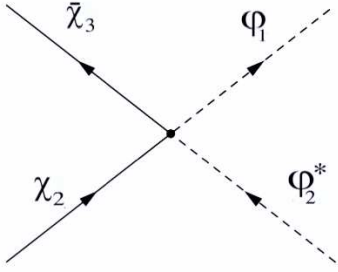}
\caption{The Feynman diagram for the effective coupling between fermion zero modes 
and ${\mathbb C}P^2$ Nambu-Goldstone bosons.
}
\label{Fig:holes}
\end{center}
\end{figure}


The interaction between the Majorana fermions and Nambu-Goldstone modes obtained in the above 
implies the instability of these modes. 
For instance, two Majorana fermions can decay into 
two Nambu-Goldstone bosons, 
as shown in Fig.~\ref{Fig:holes}.
We have not yet considered the strange quark mass.
If the effect of the strange quark mass is taken into account,
the effective potential of the ${\mathbb C}P^2$ modes is present \cite{Eto:2009tr}, which will suppress such a decay 
below the energy scale corresponding to the strange quark mass.

 \section{Summary and discussion}\label{sec:summary}
 In this paper we have derived the (1+1)-dimensional effective action of the Majorana fermion zero modes coupling to the ${\mathbb C}P^2$ Nambu-Goldstone bosonic modes 
localized around a non-Abelian vortex 
in the CFL phase of dense QCD, by using 
the method of nonlinear realizations. These fermion zero modes arise  in the vicinity of a non-Abelian vortex, so they live around the vortex core. We have integrated the action over the vortex codimensions $(x, y)$ which leaves the zero modes as the main degrees of freedom on two-dimensional space (world sheet).  The ${\mathbb C}P^2$ Nambu-Goldstone modes are 
generated by the spontaneous breaking of  $\SU(3)_{\c+\F}$ color-flavor symmetry of the CFL ground state to $\U(1)_{\c+\F}\times\SU(2)_{\c+\F}$. 
These modes appear inside the vortex core along with the fermion zero modes. To demonstrate  their interaction we have used the nonlinear realization method. The fermion zero modes transform as the triplet representation of the unbroken group $\SU(2)_{\c+\F}$, 
and this $\SU(2)_{\c+\F}$ also acts linearly on the $\mathbb{C}P^2$ Nambu-Goldstone modes as an isotropy group. We have used the property of the nonlinear action of the full group $\SU(3)_{\c+\F}$ to generate the interaction term between the $\mathbb{C}P^2$ Nambu-Goldstone modes and fermion zero modes  
by introducing a covariant derivative.

In this paper, we have studied only a single vortex.
Around a multiple winding vortex, 
there appear, in general, localized Dirac fermion doublets 
of $\SU(2)$
\cite{Fujiwara:2011za}
that also show non-Abelian statistics
\cite{Yasui:2012zb}.
Since the nonlinear realization method is generally 
 applied to arbitrary representation, 
it is straightforward to apply it to the doublet fermions
in such  cases.

It was shown in Refs.~\cite{Gorsky:2011hd,Eto:2011mk} 
that a quantum mechanical potential is induced 
on the ${\mathbb C}P^2$ modes once 
nonperturbative quantum effects are taken into account, 
and it implies the existence of confined monopoles
and quark-hadron duality \cite{Eto:2011mk}.
This potential may be changed or even eliminated 
if one takes into account fermion zero modes 
coupling to the ${\mathbb C}P^2$ modes 
\cite{Witten:1978bc}.
\section*{Acknowledgments}
This work is supported by the MEXT-Supported Program for the Strategic
Research Foundation at Private Universities ``Topological Science''
(Grant No.~S1511006).
The work of M.~N.~is supported in part by a Grant-in-Aid for
Scientific Research on Innovative Areas ``Topological Materials
Science'' (KAKENHI Grant No.~15H05855) and ``Nuclear Matter in Neutron
Stars Investigated by Experiments and Astronomical Observations''
(KAKENHI Grant No.~15H00841) from the the Ministry of Education,
Culture, Sports, Science (MEXT) of Japan. The work of M.~N.~is also
supported in part by the Japan Society for the Promotion of Science
(JSPS) Grant-in-Aid for Scientific Research (KAKENHI Grant
No.~25400268).


  \pagebreak
  
 

  


\begin{thebibliography}{0}

  
\bibitem{Alford:1997zt} 
  M.~G.~Alford, K.~Rajagopal and F.~Wilczek,
  ``QCD at finite baryon density: Nucleon droplets and color superconductivity,''
  Phys.\ Lett.\ B {\bf 422}, 247 (1998)
  [hep-ph/9711395];

\bibitem{Alford:1998mk} 
  M.~G.~Alford, K.~Rajagopal and F.~Wilczek,
  ``Color flavor locking and chiral symmetry breaking in high density QCD,''
  Nucl.\ Phys.\ B {\bf 537}, 443 (1999)
  [hep-ph/9804403].


\bibitem{Alford:2007xm} 
  M.~G.~Alford, A.~Schmitt, K.~Rajagopal and T.~Sch\"{a}fer,
  ``Color superconductivity in dense quark matter,''
  Rev.\ Mod.\ Phys.\  {\bf 80}, 1455 (2008)
  [arXiv:0709.4635 [hep-ph]].

\bibitem{Rajagopal:2000wf} 
  K.~Rajagopal and F.~Wilczek,
  ``The Condensed matter physics of QCD,''
  In *Shifman, M. (ed.): At the frontier of particle physics, vol. 3* 2061-2151 (World Scientific) 
  [hep-ph/0011333].


\bibitem{Alford:1999pa}
  M.~G.~Alford, J.~Berges and K.~Rajagopal,
  ``Unlocking color and flavor in superconducting strange quark matter,''
  Nucl.\ Phys.\  B {\bf 558}, 219 (1999)
  [arXiv:hep-ph/9903502].


\bibitem{Balachandran:2005ev} 
  A.~P.~Balachandran, S.~Digal and T.~Matsuura,
  ``Semi-superfluid strings in high density QCD,''
  Phys.\ Rev.\ D {\bf 73}, 074009 (2006)
  [hep-ph/0509276].

\bibitem{Nakano:2007dr} 
  E.~Nakano, M.~Nitta and T.~Matsuura,
  ``Non-Abelian strings in high density QCD: Zero modes and interactions,''
  Phys.\ Rev.\ D {\bf 78}, 045002 (2008)
  [arXiv:0708.4096 [hep-ph]].

\bibitem{Nakano:2008dc} 
  E.~Nakano, M.~Nitta and T.~Matsuura,
  ``Non-Abelian Strings in Hot or Dense QCD,''
  Prog.\ Theor.\ Phys.\ Suppl.\  {\bf 174}, 254 (2008)
  [arXiv:0805.4539 [hep-ph]].

%
\bibitem{Eto:2009kg} 
  M.~Eto and M.~Nitta,
  ``Color Magnetic Flux Tubes in Dense QCD,''
  Phys.\ Rev.\ D {\bf 80}, 125007 (2009)
  [arXiv:0907.1278 [hep-ph]].

\bibitem{Eto:2013hoa} 
  M.~Eto, Y.~Hirono, M.~Nitta and S.~Yasui,
  ``Vortices and Other Topological Solitons in Dense Quark Matter,''
  PTEP {\bf 2014}, no. 1, 012D01 (2014)
  [arXiv:1308.1535 [hep-ph]].


\bibitem{Giannakis:2001wz} 
  I.~Giannakis and H.~C.~Ren,
  ``The Ginzburg-Landau free energy functional of color superconductivity at weak coupling,''
  Phys.\ Rev.\ D {\bf 65}, 054017 (2002)
  [hep-ph/0108256].

\bibitem{Iida:2000ha} 
  K.~Iida and G.~Baym,
  ``The Superfluid phases of quark matter: Ginzburg-Landau theory and color neutrality,''
  Phys.\ Rev.\ D {\bf 63}, 074018 (2001)
  [Phys.\ Rev.\ D {\bf 66}, 059903 (2002)]
  [hep-ph/0011229].

\bibitem{Iida:2001pg} 
  K.~Iida and G.~Baym,
  ``Superfluid phases of quark matter. 2: phenomenology and sum rules,''
  Phys.\ Rev.\ D {\bf 65}, 014022 (2002)
  [hep-ph/0108149].

\bibitem{Nakano:2007dq}
  E.~Nakano, M.~Nitta and T.~Matsuura,
  ``Interactions of Non-Abelian Global Strings,''
  Phys.\ Lett.\  B {\bf 672}, 61 (2009)
  [arXiv:0708.4092 [hep-ph]].

\bibitem{Eto:2009bh} 
  M.~Eto, E.~Nakano and M.~Nitta,
  ``Effective world-sheet theory of color magnetic flux tubes in dense QCD,''
  Phys.\ Rev.\ D {\bf 80}, 125011 (2009)
  [arXiv:0908.4470 [hep-ph]].

\bibitem{Eto:2009tr} 
  M.~Eto, M.~Nitta and N.~Yamamoto,
  ``Instabilities of Non-Abelian Vortices in Dense QCD,''
  Phys.\ Rev.\ Lett.\  {\bf 104}, 161601 (2010)
  [arXiv:0912.1352 [hep-ph]].

\bibitem{Hirono:2010gq} 
  Y.~Hirono, T.~Kanazawa and M.~Nitta,
  ``Topological Interactions of Non-Abelian Vortices with Quasi-Particles in High Density QCD,''
  Phys.\ Rev.\ D {\bf 83}, 085018 (2011)
  [arXiv:1012.6042 [hep-ph]].

\bibitem{Hirono:2012ki} 
  Y.~Hirono and M.~Nitta,
  ``Anisotropic optical response of dense quark matter under rotation: Compact stars as cosmic polarizers,''
  Phys.\ Rev.\ Lett.\  {\bf 109}, 062501 (2012)
  [arXiv:1203.5059 [hep-ph]].

\bibitem{Vinci:2012mc} 
  W.~Vinci, M.~Cipriani and M.~Nitta,
  ``Spontaneous Magnetization through Non-Abelian Vortex Formation in Rotating Dense Quark Matter,''
  Phys.\ Rev.\ D {\bf 86}, 085018 (2012)
  [arXiv:1206.3535 [hep-ph]].

\bibitem{Forbes:2001gj} 
  M.~M.~Forbes and A.~R.~Zhitnitsky,
  ``Global strings in high density QCD,''
  Phys.\ Rev.\ D {\bf 65}, 085009 (2002)
  [hep-ph/0109173].

\bibitem{Alford:2016dco} 
  M.~G.~Alford, S.~K.~Mallavarapu, T.~Vachaspati and A.~Windisch,
  ``Stability of superfluid vortices in dense quark matter,''
  arXiv:1601.04656 [nucl-th].

\bibitem{Kobayashi:2013axa} 
  M.~Kobayashi, E.~Nakano and M.~Nitta,
  ``Color Magnetism in Non-Abelian Vortex Matter,''
  JHEP {\bf 1406}, 130 (2014)
  [arXiv:1311.2399 [hep-ph]].
  
\bibitem{Chatterjee:2015lbf} 
  C.~Chatterjee and M.~Nitta,
  ``Aharonov-Bohm Phase in High Density Quark Matter,''
  arXiv:1512.06603 [Phys. Rev. D (to be published)].
  


\bibitem{Yasui:2010yw} 
  S.~Yasui, K.~Itakura and M.~Nitta,
  ``Fermion structure of non-Abelian vortices in high density QCD,''
  Phys.\ Rev.\ D {\bf 81}, 105003 (2010)
  [arXiv:1001.3730 [hep-ph]].
\bibitem{Fujiwara:2011za} 
  T.~Fujiwara, T.~Fukui, M.~Nitta and S.~Yasui,
  ``Index theorem and Majorana zero modes along a non-Abelian vortex in a color superconductor,''
  Phys.\ Rev.\ D {\bf 84}, 076002 (2011)
  [arXiv:1105.2115 [hep-ph]].

\bibitem{Yasui:2010yh} 
  S.~Yasui, K.~Itakura and M.~Nitta,
  ``Majorana meets Coxeter: Non-Abelian Majorana Fermions and Non-Abelian Statistics,''
  Phys.\ Rev.\ B {\bf 83}, 134518 (2011)
  [arXiv:1010.3331 [cond-mat.mes-hall]];
  Y.~Hirono, S.~Yasui, K.~Itakura and M.~Nitta,
  ``Non-Abelian statistics of vortices with multiple Majorana fermions,''
  Phys.\ Rev.\ B {\bf 86}, 014508 (2012)
  [arXiv:1203.0173 [cond-mat.supr-con]].
  S.~Yasui, Y.~Hirono, K.~Itakura and M.~Nitta,
  ``Non-Abelian Vortices, Majorana Fermions and Non-Abelian Statistics,''
  JPS Conf.\ Proc.\  {\bf 4}, 013004 (2015).



\bibitem{Ivanov:2001}
D.~A.~Ivanov, 
``Non-Abelian Statistics of Half-Quantum Vortices in $\mathit{p}$-Wave Superconductors,''
Phys.\ Rev.\ Lett.\ {\bf 86}, 268--271 (2001)


\bibitem{Weinberg:1968de} 
  S.~Weinberg,
  ``Nonlinear realizations of chiral symmetry,''
  Phys.\ Rev.\  {\bf 166}, 1568 (1968).
\bibitem{Coleman:1969sm} 
  S.~R.~Coleman, J.~Wess and B.~Zumino,
  ``Structure of phenomenological Lagrangians. 1.,''
  Phys.\ Rev.\  {\bf 177}, 2239 (1969).
  C.~G.~Callan, Jr., S.~R.~Coleman, J.~Wess and B.~Zumino,
  ``Structure of phenomenological Lagrangians. 2.,''
  Phys.\ Rev.\  {\bf 177}, 2247 (1969).
\bibitem{Salam:1969rq} 
  A.~Salam and J.~A.~Strathdee,
  ``Nonlinear realizations. 1: The Role of Goldstone bosons,''
  Phys.\ Rev.\  {\bf 184}, 1750 (1969).
  \bibitem{Adler:book}
  S. L. Adler and R. F Dashen, ``Current Algebras and Applications to Particle Physics'', Benjamin/ Cummings, New York, 1968

\bibitem{Bando:1987br} 
  M.~Bando, T.~Kugo and K.~Yamawaki,
  ``Nonlinear Realization and Hidden Local Symmetries,''
  Phys.\ Rept.\  {\bf 164}, 217 (1988).
  
\bibitem{Gorsky:2011hd} 
  A.~Gorsky, M.~Shifman and A.~Yung,
  ``Confined Magnetic Monopoles in Dense QCD,''
  Phys.\ Rev.\ D {\bf 83}, 085027 (2011)
  [arXiv:1101.1120 [hep-ph]].

\bibitem{Eto:2011mk} 
  M.~Eto, M.~Nitta and N.~Yamamoto,
  ``Confined Monopoles Induced by Quantum Effects in Dense QCD,''
  Phys.\ Rev.\ D {\bf 83}, 085005 (2011)
  [arXiv:1101.2574 [hep-ph]].


\bibitem{Kibble:1967sv} 
  T.~W.~B.~Kibble,
  ``Symmetry breaking in nonAbelian gauge theories,''
  Phys.\ Rev.\  {\bf 155}, 1554 (1967).

\bibitem{Yasui:2012zb} 
  S.~Yasui, Y.~Hirono, K.~Itakura and M.~Nitta,
  ``Non-Abelian statistics of vortices with non-Abelian Dirac fermions,''
  Phys.\ Rev.\ E {\bf 87}, no. 5, 052142 (2013)
  [arXiv:1204.1164 [cond-mat.supr-con]];
  S.~Yasui, K.~Itakura and M.~Nitta,
  ``Dirac returns: Non-Abelian statistics of vortices with Dirac fermions,''
  Nucl.\ Phys.\ B {\bf 859}, 261 (2012)
  [arXiv:1109.2755 [cond-mat.supr-con]].

\bibitem{Witten:1978bc} 
  E.~Witten,
  ``Instantons, the Quark Model, and the 1/n Expansion,''
  Nucl.\ Phys.\ B {\bf 149}, 285 (1979).


\end{thebibliography}
\end{document}